\documentclass[%
 reprint,
superscriptaddress,
frontmatterverbose, 
 amsmath,amssymb,
 aps,
prb,
]{revtex4-2}

\usepackage{graphicx}
\usepackage{dcolumn}
\usepackage{bm}

\usepackage{color}
\usepackage[normalem]{ulem}
\usepackage[utf8]{inputenc}
\usepackage[colorlinks=true,allcolors=blue]{hyperref}
\usepackage[mathlines]{lineno}
\usepackage{braket}

\begin{document}

\preprint{APS/123-QED}

\title{Assessing carrier mobility, dopability, and defect tolerance in the chalcogenide perovskite BaZrS$_3$}

\author{Zhenkun Yuan}
\altaffiliation{These authors contributed equally}
\affiliation{Thayer School of Engineering, Dartmouth College, Hanover, New Hampshire 03755, USA}
\author{Diana Dahliah}
\altaffiliation{These authors contributed equally}
\affiliation{Institute of Condensed Matter and Nanosciences, Universit\'{e} catholique de Louvain, Chemin \'{e}toiles 8, bte L7.03.01, Louvain-la-Neuve 1348, Belgium}
\affiliation{Department of Physics, An-Najah National University, Nablus, Palestine}
\author{Romain Claes}\affiliation{Institute of Condensed Matter and Nanosciences, Universit\'{e} catholique de Louvain, Chemin \'{e}toiles 8, bte L7.03.01, Louvain-la-Neuve 1348, Belgium}
\affiliation{School of Chemistry, University of Birmingham, Edgbaston, Birmingham, B15 2TT, UK}
\author{Andrew Pike}
\affiliation{Thayer School of Engineering, Dartmouth College, Hanover, New Hampshire 03755, USA}
\author{David P. Fenning}
\affiliation{Department of Nanoengineering, University of California, San Diego, La Jolla, CA 92093, USA}
\author{Gian-Marco Rignanese}
\affiliation{Institute of Condensed Matter and Nanosciences, Universit\'{e} catholique de Louvain, Chemin \'{e}toiles 8, bte L7.03.01, Louvain-la-Neuve 1348, Belgium}
\author{Geoffroy Hautier}
\email{geoffroy.hautier@dartmouth.edu}
\affiliation{Thayer School of Engineering, Dartmouth College, Hanover, New Hampshire 03755, USA}

\date{\today}

\begin{abstract}
The chalcogenide perovskite BaZrS$_3$ has attracted much attention as a promising solar absorber for thin-film photovoltaics. Here, we use first-principles calculations to evaluate its carrier transport and defect properties. We find that BaZrS$_3$ has a phonon-limited electron mobility of 37 cm$^2$/Vs comparable to that in halide perovskites but lower hole mobility of 11 cm$^2$/Vs. The defect computations indicate that BaZrS$_3$ is intrinsically n-type due to shallow sulfur vacancies, but that strong compensation by sulfur vacancies will prevent attempts to make it p-type. We also establish that BaZrS$_3$ shows some degree of defect tolerance, presenting only few low formation energy, deep intrinsic defects. Among the deep defects, sulfur interstitials are the dominant nonradiative recombination centers but exhibit a moderate capture coefficient. Our work highlights the material's intrinsic limitations in carrier mobility and p-type doping and suggests focusing on suppressing the formation of sulfur interstitials to reach longer carrier lifetime.
\end{abstract}

\maketitle

\section{\label{sec:level1}INTRODUCTION}

Lead halide perovskites have revolutionized the field of photovoltaics (PV) by opening a promising path to earth-abundant, easily processable, and high-efficiency thin-film technologies \cite{kojima2009organometal,nrelchart}. The exceptional PV performance of halide perovskites is however overshadowed by their poor stability \cite{correa2017promises,park2019intrinsic,li2024fundamental}. Structural analogy has motivated the search for alternative solar absorbers forming in the perovskite structure but in chemistries that could be more stable \cite{sun2015chalcogenide,korbel2016stability,zhao2017design,kuhar2017sulfide,chakraborty2017rational,lu2018accelerated,huo2018high,cai2019high,gebhardt2023screening,wu2023tilt}. The chalcogenide perovskites ABX$_3$ (A=Ca, Sr, Ba, B=Ti, Zr, Hf, and X=S, Se) have emerged in this context with their first suggestion as solar absorbers coming from first-principles studies \cite{sun2015chalcogenide} followed by experimental synthesis and characterization especially of BaZrS$_3$ \cite{sopiha2022chalcogenide,choi2022emerging,meng2016alloying,perera2016chalcogenide,niu2017bandgap}. BaZrS$_3$ shows excellent stability in ambient conditions and exhibits a $\sim$1.8 eV direct band gap which can be tuned to 1.5 eV by alloying with BaTiS$_3$ or BaZrSe$_3$ \cite{sopiha2022chalcogenide,meng2016alloying,perera2016chalcogenide,niu2017bandgap,wei2020ti,comparotto2020chalcogenide,sadeghi2023expanding,ye2024processing,kayastha2024first,bystricky2024thermal}. Significant efforts have been dedicated to growing high-quality thin films of BaZrS$_3$ and its alloys, using a range of techniques such as pulsed laser deposition \cite{wei2020realization,YU2021105959,ramanandan2023understanding,surendran2024hybrid}, sputtering \cite{comparotto2020chalcogenide,HAN2023145351,jamshaid2024synthesis}, molecular beam epitaxy \cite{sadeghi2021making,surendran2021epitaxial}, and solution-based synthesis \cite{ravi2021colloidal,turnley2022solution,pradhan2023synthesis}. Very recently, a proof-of-concept BaZrS$_3$ solar cell has been reported, demonstrating an efficiency of 0.11\% \cite{dallas2024exploring}. Interestingly, BaZrS$_3$ also stands out as a top candidate in a few high-throughput computational screening of thin-film solar absorbers \cite{yuan2024discovery,kuhar2018,huo2018high,fabini2019candidate}.

In this work, we use first-principles calculations to clarify the carrier transport and defect properties in BaZrS$_3$. We compute the phonon-limited carrier mobility showing that BaZrS$_3$ has intrinsically low hole mobility. We also perform state-of-the-art hybrid-functional defect calculations. We show that BaZrS$_3$ is intrinsically n-type, and that p-type doping of BaZrS$_3$ will be very difficult due to strong compensation by intrinsic donor defects. We identify the sulfur interstitial (S$_i$) as the main deep defect, but show that its nonradiative capture coefficient is moderate and that sulfur chemical potential control should mitigate its impact on the carrier lifetime in BaZrS$_3$. Our results suggest pathways regarding growth condition optimization and device design towards high-performance BaZrS$_3$ absorbers.

\section{RESULTS}

\subsection{\label{sec:level2}Electronic band structure and carrier transport}

\begin{figure*}[bht]
    \centering
    \includegraphics[width=0.88\textwidth]{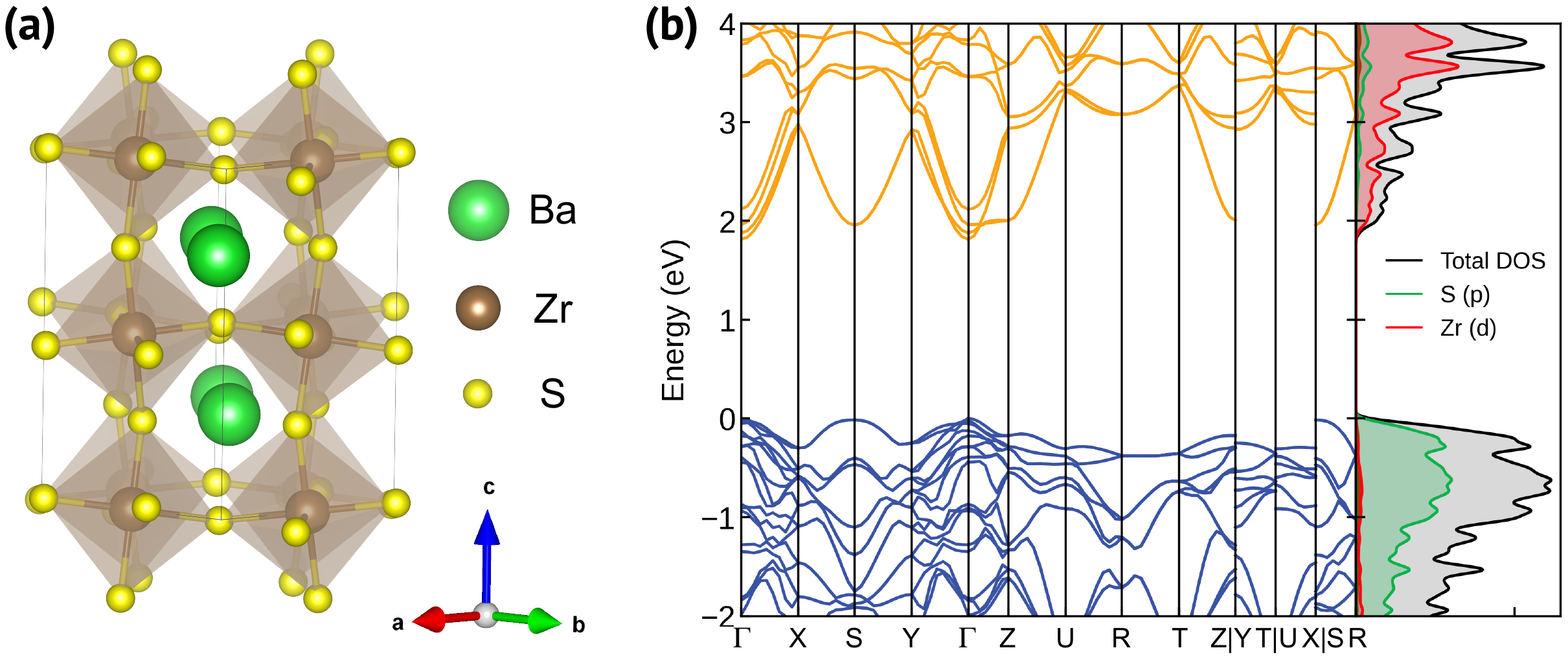}
    \caption{(a) Crystal structure of BaZrS$_3$.
    (b) HSE06-calculated electronic band structure and (partial) density of states (DOS) of BaZrS$_3$.}
    \label{fig1}
\end{figure*}

BaZrS$_3$ forms in an orthorhombic \textit{Pnma} perovskite structure [Fig.~\ref{fig1}(a)]. Our calculated band structure using the Heyd-Scuseria-Ernzerhof hybrid functional (HSE06) \cite{Heyd2003} shows a direct band gap of 1.81 eV at the $\Gamma$ point [Fig.~\ref{fig1}(b)], in agreement with previous calculations and experiment \cite{Polfus2015,sun2015chalcogenide,huo2018high,meng2016alloying,perera2016chalcogenide,niu2017bandgap,wei2020ti,comparotto2020chalcogenide,sadeghi2023expanding,ye2024processing}. Combining computed electronic and phonon properties, we find a phonon-limited carrier mobility of 11 cm$^2$/Vs for holes and 37 cm$^2$/Vs for electrons at room temperature (see Sec.~\ref{section_ma} for details), with the carrier scattering mechanism dominated by polar optical phonons. These values are upper bounds as realistic polycrystalline films will have additional scatterings from grain boundaries, impurities, and others. They are much lower than those calculated for conventional thin-film inorganic solar absorbers (such as CdTe \cite{ganose2021efficient,Vukmirovi2021} and Cu$_2$ZnSnS$_4$ \cite{Monserrat2018}). The calculated hole mobility of BaZrS$_3$ is also lower than for the halide perovskite CH$_3$NH$_3$PbI$_3$ (11 vs 47 cm$^2$/Vs), yet these two materials have comparable calculated electron mobility \cite{ponce2019origin}. Our results are consistent with experiments on BaZrS$_3$ thin films which indicate low carrier mobilities ($\sim$2 cm$^2$/Vs for holes and $\sim$10–20 cm$^2$/Vs for electrons) \cite{wei2020realization,Mrquez2021,YU2021105959,ravi2021colloidal}. The measured low carrier mobility has been often attributed to small grain size or impurity scattering \cite{wei2020realization,YU2021105959}. While these could be limiting factors in the experiments, our results highlight that BaZrS$_3$ has intrinsically low phonon-limited carrier mobility, and that experimentally it is very unlikely to achieve mobilities higher than our computed values. We note that Ye \textit{et al}. reported a very high sum mobility ($>$100 cm$^2$/Vs) in BaZrS$_3$ films based on time-resolved photoluminescence (TRPL) measurements but the data suffer from reported very large uncertainty \cite{ye2022time}.

The large difference between hole and electron mobilities directly comes from the electronic band structure. The lower conduction bands are much more dispersive than the upper valence bands [Fig.~\ref{fig1}(b)]. As a result, the effective mass is found to be small for electrons (0.3 $m_0$) and relatively large for holes (0.9 $m_0$). The fundamental difference in hole effective mass and mobility between BaZrS$_3$ and CH$_3$NH$_3$PbI$_3$ comes from the different electronic character in the valence band. While the halide perovskite mixes anion and cation orbitals leading to delocalized valence band \cite{yin2015halide}, the sulfide shows a more ionic behavior with the valence band being mainly of anion character [Fig.~\ref{fig1}(b)].

\subsection{Intrinsic point defects and doping}

We have calculated all the intrinsic point defects in BaZrS$_3$ including the vacancies ($V_\mathrm{Ba}$, $V_\mathrm{Zr}$, $V_\mathrm{S}$), interstitials (Ba$_i$, Zr$_i$, S$_i$), and antisites (Ba$_\mathrm{Zr}$, Zr$_\mathrm{Ba}$, Ba$_\mathrm{S}$, S$_\mathrm{Ba}$, Zr$_\mathrm{S}$, S$_\mathrm{Zr}$). Our first-principles calculations are all performed using the HSE06 hybrid functional and large $3\times3\times2$ supercell (360 atoms) with appropriate charge correction and spin-polarization, which is different from previous first-principles calculations \cite{meng2016alloying,wu2021defect}. We provide in Sec.~\ref{section_mb} the details of our methodology and a comparison to previous calculations in Supplemental Material \cite{supp}. We note that some of the defects (e.g., S$_i$) involve several configurations that are close in energy. In the following, we report only results for the lowest energy configurations, while those for metastable configurations can be found in Fig. S2 of the Supplemental Material \cite{supp}. 

\begin{figure*}[bht]
    \centering
    \includegraphics[width=1\textwidth]{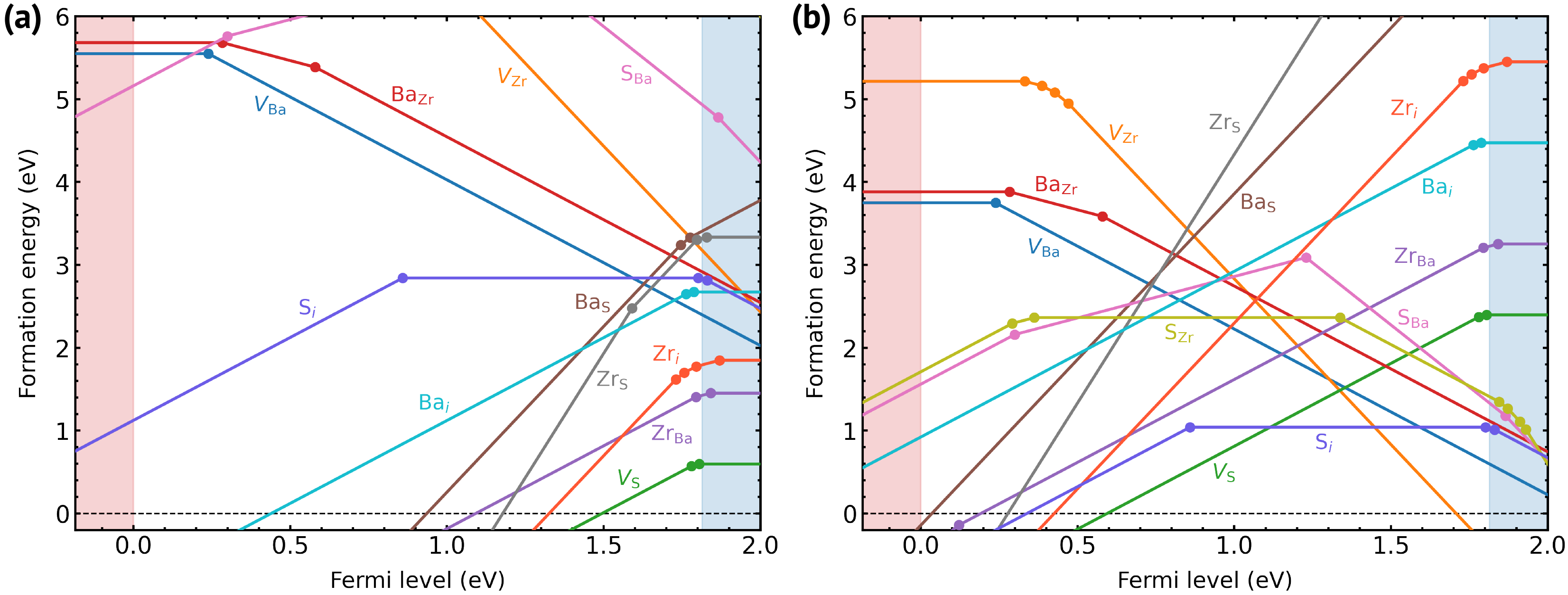}
    \caption{Formation energies of intrinsic point defects in BaZrS$_{3}$ as a functional of Fermi level, under (a) S-poor and (b) S-rich conditions. The Fermi level is referenced to the valence-band maximum (VBM) of BaZrS$_3$. The slopes of the formation-energy lines indicate defect charge states, and the dots denote charge-state transition levels (see also Fig.~\ref{fig3}).}
    \label{fig2}
\end{figure*}

\begin{figure*}[bht]
    \centering
    \includegraphics[width=0.8\textwidth]{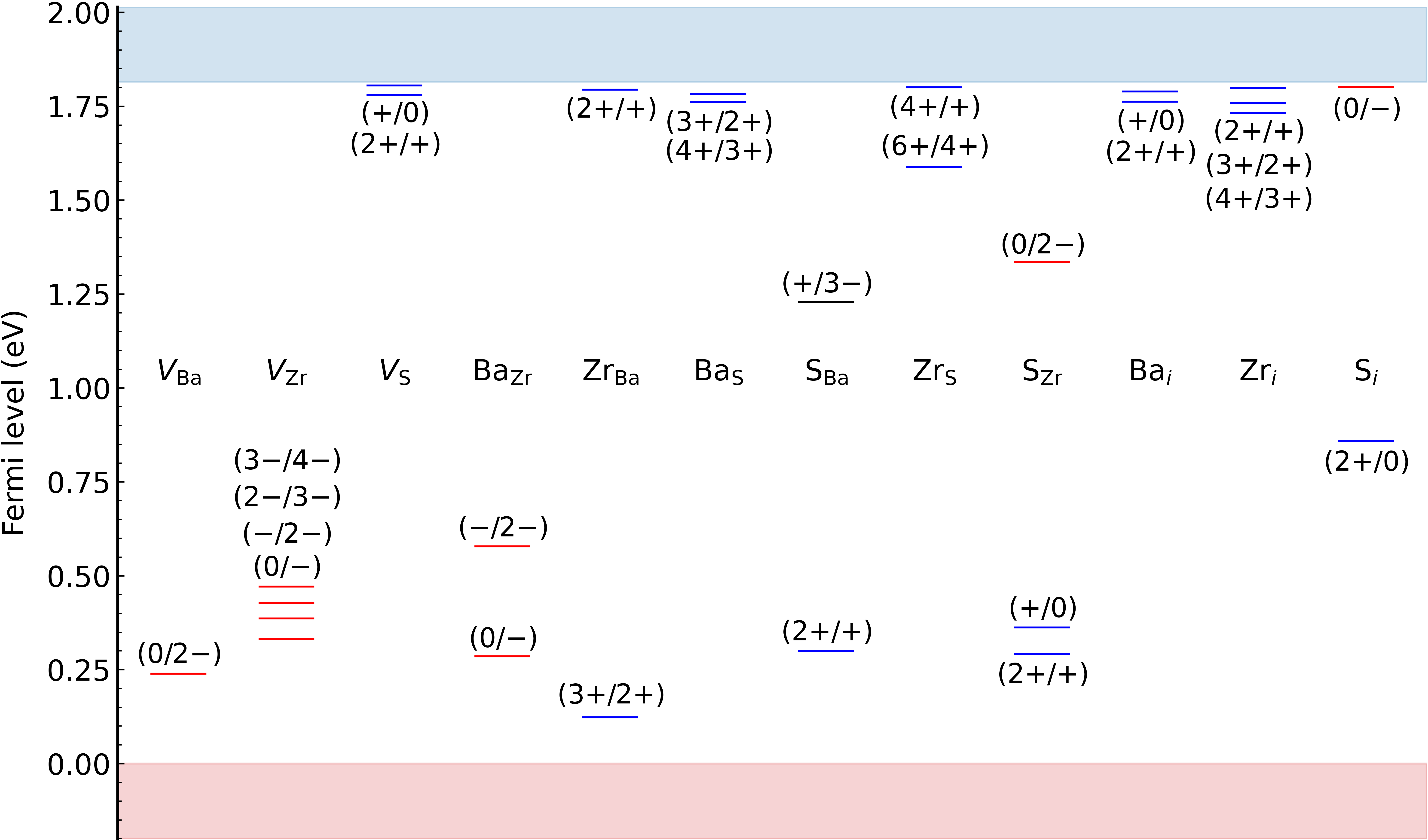}
    \caption{Defect charge-state transition levels $\epsilon(q/q^\prime)$. Only the levels falling into the BaZrS$_3$ band gap are shown. The red and blue bars denote acceptor and donor levels, respectively.}
    \label{fig3}
\end{figure*}

Fig.~\ref{fig2} shows the formation energies of the intrinsic defects in BaZrS$_3$ for both S-poor and S-rich conditions (See Fig. S1 in the Supplemental Material for details \cite{supp}). The defect charge-state transition levels are plotted in Fig.~\ref{fig3}. We find that a series of shallow donor defects can form in BaZrS$_3$. The $V_\mathrm{S}$ is the dominant donor, giving rise to two donor levels ($+$/0) and (2$+$/$+$) that are almost in resonance with the conduction band. Under S-poor conditions [Fig.~\ref{fig2}(a)], the $V_\mathrm{S}$ has fairly low formation energy and thus exists in high concentration. On the other hand, the acceptor defects, mainly $V_\mathrm{Ba}$ and Ba$_\mathrm{Zr}$, have high formation energies. These indicate that S-poor BaZrS$_3$ will be heavily n-type doped by the $V_\mathrm{S}$ donors. Under S-rich conditions [Fig.~\ref{fig2}(b)], the formation energy of $V_\mathrm{S}$ is increased, while the formation energies of the acceptor defects are reduced. Under those conditions, the equilibrium Fermi level would be pinned close to the intersection of the formation energies of $V_\mathrm{S}$ and $V_\mathrm{Zr}$ (about 0.5 eV below the conduction band), indicative of a very weak n-type, almost intrinsic BaZrS$_3$. Our results explain the experimental observation that as-grown BaZrS$_3$ is intrinsically n-type with the electron concentration as high as 10$^{19}$–10$^{20}$ cm$^{-3}$ \cite{wei2020realization}, and we attribute this doping to sulfur vacancies. Very recently, Aggarwal \textit{et al}.  showed that BaZrS$_3$ films synthesized in a S-rich atmosphere are insulating but become n-type conductive with subsequent annealing in high vacuum (i.e., a S-poor environment) \cite{aggarwal2024charge}. They attributed this transition to the formation of sulfur vacancies. Additionally, they found no significant change in conductivity with varying Ba/Zr ratio. These findings agree with our calculations.

Fig.~\ref{fig2} also indicates that it will be very difficult to achieve p-type BaZrS$_3$. While $V_\mathrm{Ba}$, $V_\mathrm{Zr}$, and Ba$_\mathrm{Zr}$ are shallow acceptors, they are strongly compensated by the $V_\mathrm{S}$ donors. Even under the most favorable S-rich conditions, Fermi-level pinning energy for p-type doping is 0.6 eV above the VBM, caused by the $V_\mathrm{S}$ whose formation energy drops first to zero when the Fermi level is approaching the VBM [Fig.~\ref{fig2}(b)]. In view of the high p-type pinning limit, any extrinsic shallow acceptors will be strongly compensated, thus preventing p-type doping \cite{robertson2011limits,walsh2017instilling,yuan2024first}. In the literature, p-type BaZrS$_3$ has only been reported once with hole concentration of $\sim$10$^{18}~\mathrm{cm}^{-3}$ \cite{HAN2023145351}. We note that this was achieved in a sample which is extremely Ba-deficient (Ba/Zr ratio as low as  $\sim$0.6), raising questions about possible secondary phases \cite{vincent2024expanding}.

Next to doping, from Figs.~\ref{fig2} and \ref{fig3} we identify a few defects with deep transition levels, including $V_\mathrm{Zr}$ (3$-$/4$-$), Ba$_\mathrm{Zr}$ ($-$/2$-$), S$_\mathrm{Ba}$ ($+$/3$-$), S$_\mathrm{Zr}$ (0/2$-$), and S$_i$ (0/2$+$). Only S$_i$ has sufficiently low formation energy to exist in significant concentrations (when in S-rich conditions; see Table S4 in the Supplemental Material for the calculated defect concentrations \cite{supp}). Since defect-assisted nonradiative recombination is one of the key processes that limit carrier lifetime and ultimately solar cell performance \cite{repins2013indications,brandt2015identifying,jaramillo2016transient,park2018point,guillemoles2019guide,kim2020upper,wang2024upper}, it is necessary to assess whether the S$_i$ is an efficient nonradiative carrier recombination center. We note that the charge-compensated Frenkel pair $[V_\mathrm{S}$+S$_i]$ might form in BaZrS$_3$, considering that the S$_i$ can also occur in negative charge states (i.e., acceptor-like behavior). However, the binding energy of this defect pair is found to be relatively small (0.74 eV; see Sec.~\ref{section_mb}), suggesting that it will not be very stable in the high-temperature processing of BaZrS$_3$.

\subsection{Nonradiative carrier capture by sulfur interstitials}

\begin{figure*}[bht]
    \centering
    \includegraphics[width=1\textwidth]{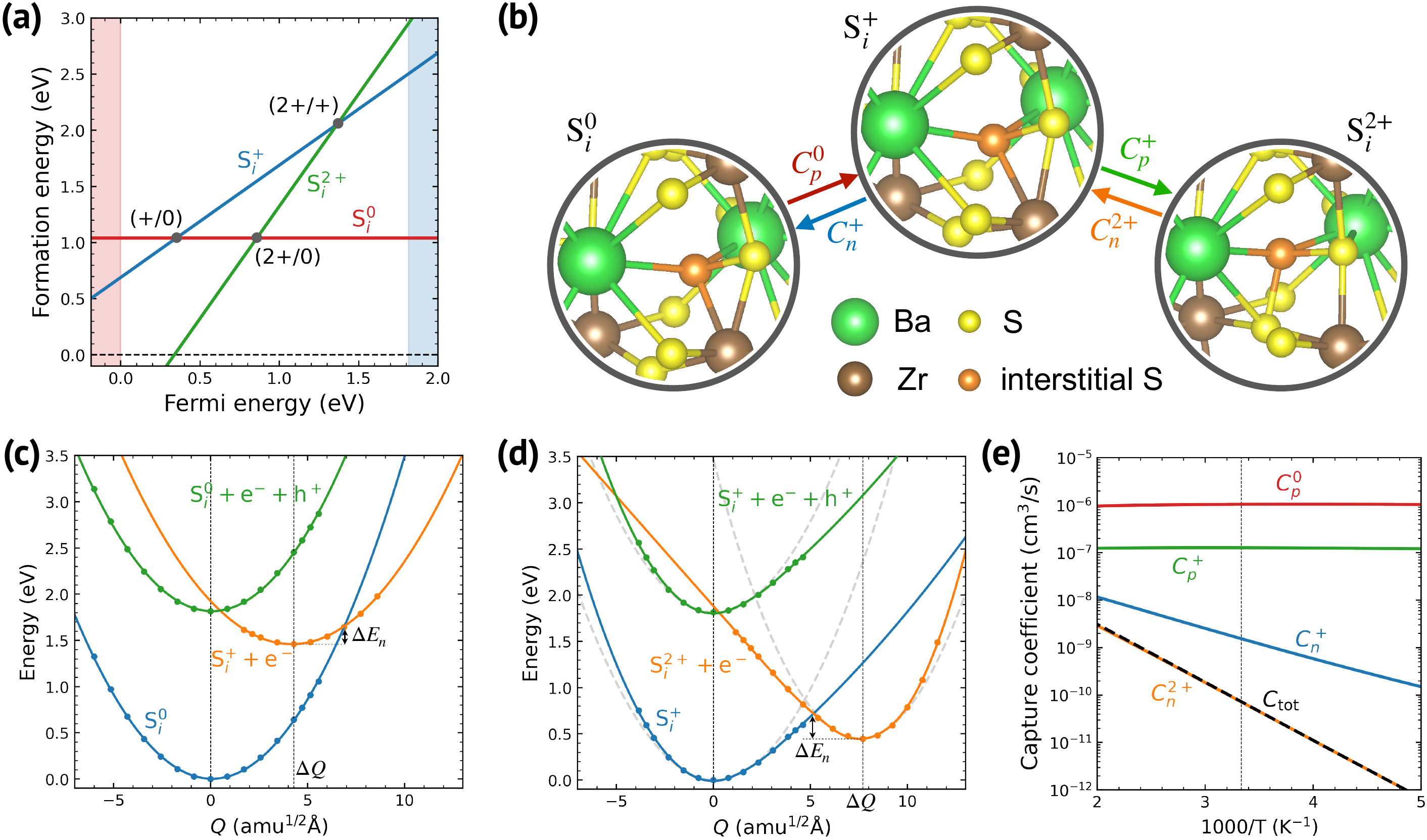}
    \caption{(a) Formation energy of the sulfur interstitial (S$_i$) as a function of Fermi level under S-rich conditions. (b) Local atomic structures of the S$_i$ in the 0, $+$, and 2$+$ charge states. (c)–(d) Configuration coordinate diagrams for the S$^0_i$ $\rightleftharpoons$ S$^+_i$ and S$^+_i$ $\rightleftharpoons$ S$^{2+}_i$ transitions. The data points are potential energies obtained from first-principles calculations. In (c), the solid colored curves are parabolic fits to the data points and thus depict the potential energy surfaces in the harmonic approximation. In (d), the solid colored curves are potential energy surfaces obtained from quadratic spline fits to the data points, and the gray dashed curves are parabolic fits. $\Delta Q$ denotes the structural difference in configuration coordinate between the initial and final charge states, and $\Delta E_n$ the electron capture barrier. (e) Temperature-dependent nonradiative capture coefficients of the S$_i$.}
    \label{fig4}
\end{figure*}

We now compute the nonradiative carrier capture coefficients of the S$_i$ (see Sec.~\ref{section_mc} for computational details). For carrier capture by the S$_i$, the relevant charge-state transitions are ($+$/0) and (2$+$/$+$), which are located at 0.35 and 1.37 eV above the VBM, respectively, as shown in Fig.~\ref{fig4}(a). There are two capture processes for electron-hole recombination via the ($+$/0) level: $C_p^0$ for hole ($p$) capture and $C_n^+$ for electron ($n$) capture, where the superscript denotes the initial charge state. Similarly, recombination via the (2$+$/$+$) level involves two capture processes: $C_p^+$ and $C_n^{2+}$. 

To illustrate the capture processes, Fig.~\ref{fig4}(b) shows the local atomic structures of the S$_i$ in the three charge states: 0, $+$, and 2$+$. The local structures of S$^0_i$ and S$^+_i$ are similar but differ from that of S$^{2+}_i$. In the 0 and $+$ charge states, the interstitial S forms distorted tetrahedral bonds with two Ba, one Zr, and one S nearest neighbors. With the S$^0_i$ $\rightarrow$  S$^+_i$ transition (i.e., capturing a hole, $C_p^0$), another lattice S atom moves towards the interstitial S. This lattice S atom and the interstitial S move further closer with the S$^+_i$  $\rightarrow$  S$^{2+}_i$ transition (i.e., capturing another hole, $C_p^+$). As a result, the S$^{2+}_i$ forms a S trimer.

The configuration coordinate diagrams for the S$^0_i$ $\rightleftharpoons$ S$^+_i$ and S$^+_i$ $\rightleftharpoons$ S$^{2+}_i$ transitions are shown in Figs.~\ref{fig4}(c) and~\ref{fig4}(d). Such diagrams map the potential energy surfaces of a defect in two adjacent charge states for a given transition as a function of a generalized configuration coordinate (Q) \cite{lang1975nonradiative,huang1981lattice,alkauskas2014first}. We find that the displacement $\Delta Q$ is larger between S$^+_i$ and S$^{2+}_i$ than between S$^0_i$ and S$^+_i$, reflecting the structural differences discussed above. The configuration coordinate diagrams indicate anharmonic atomic vibrations in the S$_i$ capture processes, specifically for the S$^+_i$ $\rightleftharpoons$ S$^{2+}_i$ transitions, as shown in Fig.~\ref{fig4}(d) (see also Fig. S3 in the Supplemental Material \cite{supp}). Anharmonic potential energy surfaces have been widely found for nonradiative carrier capture in halide perovskites and other low-symmetry semiconductors \cite{Zhang2020,whalley2021giant,zhang2022defect,whalley2023,kumagai2023alkali,huang2023metastability,zhang2023iodine,wang2024upper}. As seen in Fig.~\ref{fig4}(d), the anharmonicity substantially lowers the capture barriers as compared to the harmonic case; this leads to vanishingly small hole capture barrier for S$^+_i$ and a moderate electron capture barrier of 0.27 eV for S$^{2+}_i$. In the case of S$^0_i$ $\rightleftharpoons$ S$^+_i$ transitions, the hole capture barrier for S$^0_i$ is also nearly zero, and electron capture by S$^+_i$ has an energy barrier of 0.17 eV [Fig.~\ref{fig4}(c)]. From the calculated capture barriers, we expect fast hole capture and slower electron capture by the S$_i$.

Fig.~\ref{fig4}(e) shows the calculated capture coefficients of the S$_i$ as a function of temperature. Consistent with the negligibly small hole capture barriers and the finite electron capture barriers, the capture coefficients indicate fast hole capture and much slower electron capture by the S$_i$. Throughout the temperature range considered (200–500 K), the hole capture coefficient of S$^0_i$ ($C_p^0$) is as high as $10^{-6}~\mathrm{cm}^{3}/\mathrm{s}$, and that of S$^+_i$ ($C_p^+$) is on the order of $10^{-7}~\mathrm{cm}^{3}/\mathrm{s}$. In contrast, at 300 K, the electron capture coefficient of S$^+_i$ ($C_n^+$) is $1.54\times10^{-9}~\mathrm{cm}^{3}/\mathrm{s}$, and that of S$^{2+}_i$ ($C_n^{2+}$) has a moderate value of $7.18\times10^{-11}~\mathrm{cm}^{3}/\mathrm{s}$. For low-doped or intrinsic BaZrS$_3$, by balancing electron and hole capture under steady-state conditions, the total capture coefficient ($C_\mathrm{tot}$) of the S$_i$ is given by \cite{alkauskas2016role,huang2023metastability}, 
\begin{equation*}
    C_\mathrm{tot} = \frac{C_n^+ + C_p^+}{1+\frac{C_n^+}{C_p^0}+\frac{C_p^+}{C_n^{2+}}}.
\end{equation*}
The $C_\mathrm{tot}$ of the S$_i$ is calculated to be $7.27\times10^{-11}~\mathrm{cm}^{3}/\mathrm{s}$ at 300 K [Fig.~\ref{fig4}(e)]. The $C_\mathrm{tot}$ is limited by $C_n^{2+}$. The capture coefficient is moderate and is two orders of magnitude smaller than the value ($7\times10^{-9}~\mathrm{cm}^{3}/\mathrm{s}$) for the dominant recombination centers (iodine interstitials, I$_i$) in $\mathrm{CH}_3\mathrm{NH}_3\mathrm{Pb}\mathrm{I}_3$ computed in a similar theoretical framework \cite{Zhang2020,zhang2022defect}. Yet, the capture coefficient of the S$_i$ is not negligible. A high density of S$_i$ can still cause strong nonradiative recombination and limit the carrier lifetime.

\section{Discussion}

BaZrS$_3$ shows higher electron mobility than hole mobility which would suggest using this material as a p-type absorber layer as it is the diffusion length of minority carriers that mainly controls the efficiency of p-n junction solar cells \cite{kitai2018principles,gokmen2013minority,scarpulla2023cdte}. Our analysis however shows that it will be unlikely to make p-type BaZrS$_3$. Using the intrinsically n-type doped BaZrS$_3$ as an absorber layer will lead to smaller minority-carrier diffusion lengths limited by the lower hole mobility and also cause issues at the device level as discussed for other n-type thin-film absorbers \cite{hill2023widegap,javaid2018thin,gershon2016photovoltaic}. We thus suggest that it could be more viable to devise a p-i-n (or n-i-p) cell using BaZrS$_3$ as the intrinsic layer (lightly n-type doped), as in halide perovskite solar cells \cite{shao2023recent}. In such a p-i-n device, the carrier diffusion length is controlled by the ambipolar mobility \cite{akel2023relevance,wang2021impact,shin2022high} which is estimated to be 17 cm$^2$/Vs using our calculated intrinsic electron and hole mobilities. Preparing intrinsic BaZrS$_3$ requires to reduce dramatically the concentration of $V_\mathrm{S}$ donors which can be achieved using S-rich growth conditions \cite{aggarwal2024charge}. However, S-rich conditions would enhance the formation of sulfur interstitials which we have shown to be the main nonradiative recombination centers. It might be beneficial to reduce electron concentration by introducing an extrinsic shallow acceptor while keeping a low sulfur chemical potential.

Since high-quality BaZrS$_3$ films are typically deposited at temperatures of $\sim900~^{\circ}$C or higher \cite{wei2020realization,comparotto2020chalcogenide,Mrquez2021,ramanandan2023understanding}, we estimate now a realistic upper bound of the S$_i$ density in high-temperature synthesized BaZrS$_3$ samples. Under S-rich conditions [which corresponds to the defect formation energies displayed in Fig.~\ref{fig2}(b)] and assuming 1000 K growth of BaZrS$_3$ and rapid quenching to room temperature, the S$_i$ density is estimated to be $4.14\times10^{17}~\mathrm{cm}^{-3}$ (see Table S4). This high S$_i$ density leads to a nonradiative recombination coefficient ($A$) of $3.01\times10^7~\mathrm{s}^{-1}$ at room temperature; here $A$ is defined as $A=N_\mathrm{d}C_\mathrm{tot}$, where $N_\mathrm{d}$ is the defect density \cite{li2019effective,das2020deep,zhang2022defect}. As a result, the nonradiative lifetime ($\tau=1/A$) is of 33 ns for S-rich conditions. Moving to S-poor conditions and/or lower synthesis temperature will reduce the S$_i$ density and increase the carrier lifetime. We note though that growing BaZrS$_3$ at moderate temperatures tends to require S-rich conditions, including recent reports of synthesis of BaZrS$_3$ films at about $600~^{\circ}$C in the presence of excess sulfur \cite{wang2000synthesis,comparotto2022synthesis,aggarwal2024charge}. Our results are in reasonable quantitative agreement with the carrier lifetime of about 50 ns measured by TRPL on BaZrS$_3$ single-crystal samples \cite{ye2022time, zhao2024photoconductive}. In comparison, the I$_i$ in $\mathrm{CH}_3\mathrm{NH}_3\mathrm{Pb}\mathrm{I}_3$ has been found to lead to a nonradiative lifetime on the order of 100 ns \cite{Zhang2020,zhang2022defect}, based on the fact that the deep-level trap density in solution-processed $\mathrm{CH}_3\mathrm{NH}_3\mathrm{Pb}\mathrm{I}_3$ samples is on the order of $10^{15}~\mathrm{cm}^{-3}$ \cite{Baumann2015,heo2017deep,yang2017iodide}.

Ye \textit{et al}. \cite{ye2022time} estimated the solar cell figure of merit ($F_\mathrm{PV}$) of BaZrS$_3$ based on experimental data and using $F_\mathrm{PV}=\alpha * L_\mathrm{D}$, where $\alpha$ is the optical absorption coefficient and $L_\mathrm{D}=\sqrt{\frac{\mu k_\mathrm{B}T}{e}\tau}$ the carrier diffusion length \cite{hodes2015understanding,akel2023relevance,crovetto2024figure,whalley2017perspective}. They found a $F_\mathrm{PV}$ value of 2.1 using an absorption coefficient of 4940 cm$^{-1}$, nonradiative lifetime of 50 ns, and ambipolar mobility of 146.2 cm$^2$/Vs \cite{ye2022time}. Our computational results do not disagree with the carrier lifetime but raise strong doubts on the mobility value. Since our calculated ambipolar mobility is about an order of magnitude lower, we estimate the $F_\mathrm{PV}$ to be just 0.7. Our results however indicate that if the S$_i$ concentration is lowered or the interstitials are passivated in some way, higher carrier lifetime could be reached which will boost the figure of merit. 

In addition to intrinsic defects, we briefly mention that BaZrS$_3$ is tolerant to oxygen impurities which could be present in high concentrations in the samples prepared by sulfurization of BaZrO$_3$ precursor \cite{wei2020realization,Pandey2020}. We find that the oxygen-related point defects, including oxygen interstitial (O$_i$) and O substitution on the S site (O$_\mathrm{S}$), are electrically inactive, i.e., they are stable in the neutral charge state for almost the entire range of Fermi levels (see Fig. S4 of the Supplemental Material \cite{supp}), in agreement with previous experimental and theoretical studies \cite{Pandey2020}.

\section{Conclusions}

We evaluated carrier transport and defect properties in the chalcogenide perovskite solar absorber BaZrS$_3$. Our results show that BaZrS$_3$ has a lower hole mobility than electron mobility (11 vs 37 cm$^2$/Vs). The mobility in this sulfide perovskite is lower and more asymmetric (in terms of hole versus electron mobilities) than in lead halide perovskites. Our defect computations indicate an intrinsic tendency for n-type doping due to the shallow donor $V_\mathrm{S}$ and that p-type doping is very unlikely to be achievable. BaZrS$_3$ shows favorable intrinsic defect behavior with only few deep defects that could act as nonradiative recombination centers. Our computations further reveal that the most problematic deep defect, the S$_i$, exhibits only a moderate nonradiative capture coefficient ($7.27\times10^{-11}~\mathrm{cm}^{3}/\mathrm{s}$). Under S-rich conditions, the carrier capture by S$_i$ will lead to a computed carrier lifetime of around 30 ns. BaZrS$_3$ exhibits some degree of defect tolerance which is beneficial for its use as solar absorbers, and we suggest that suppressing the formation of S$_i$ will be critical in further increasing the carrier lifetime in BaZrS$_3$.

\section{Methods}

\subsection{Carrier mobility calculations\label{section_ma}}

The phonon-limited carrier mobility was calculated using AMSET (Ab initio Scattering and Transport). AMSET is a Python package for performing efficient first-principles calculations of carrier scattering rates and transport in solid-state semiconductors and insulators \cite{ganose2021}. It can account for different carrier scattering mechanisms, e.g., electron-phonon coupling. To prepare inputs for AMSET calculations, we calculated the material parameters of BaZrS$_3$ using the VASP code and PBEsol functional \cite{kresse1996efficient,kresse1999ultrasoft,perdew2008restoring}. A \textbf{k}-point grid of 8$\times$8$\times$6 and an energy cut-off of 400 eV were found to converge the total energy of the \textit{Pnma} unit-cell structure of BaZrS$_3$ to within 1 meV/atom. In this work, only carrier scattering from polar optical phonons and acoustic deformation potentials were considered. For AMSET calculation of the carrier scattering rates, a dense \textbf{k}-point grid of 18$\times$18$\times$12 was employed to obtain the single-electron wave functions which were used for calculation of the electron-phonon matrix elements from density-functional perturbation theory \cite{baroni1987elastic,giustino2017electron}. The calculated material parameters and full mobility tensor are detailed in the Supplemental Material \cite{supp}, with a discussion on possible errors.

\subsection{Defect calculations\label{section_mb}}

The defect calculations were performed using the VASP code and the screened hybrid density functional of Heyd–Scuseria–Ernzerhof (HSE06) \cite{kresse1996efficient,kresse1999ultrasoft,heyd2003hybrid}. The standard VASP projector augmented wave (PAW) pseudopotentials (version PBE$\_$54) were used: Ba\_sv (10 valence electrons), Zr\_sv (12 valence electrons), and S (6 valence electrons). A plane-wave basis set with a cutoff of 400 eV was used for expanding electron wave functions. Unless explicitly mentioned, these settings were employed consistently in all our defect calculations. 

The defect calculations started from relaxation of the BaZrS$_3$ orthorhombic \textit{Pnma} unit cell. Using a 6$\times$6$\times$4 $\Gamma$-centered \textbf{k}-point mesh, the HSE06-calculated lattice constants are: $a=7.023$ \AA, $b=7.119$ \AA, and $c=10.006$ \AA, in good agreement with previous HSE06 calculations and experiments \cite{meng2016alloying,lelieveld1980sulphides}. Based on the HSE06-relaxed unit cell, a 3$\times$3$\times$2 supercell which contains 360 atoms was constructed and used for modeling point defects. For the supercell containing a point defect, all the internal atomic positions were fully relaxed with a force convergence criterion of 0.01 eV/\AA. A $\Gamma$-only \textbf{k}-point mesh was used for the supercell calculations. Spin polarization was explicitly considered in all the defect calculations.

Defect formation energies were computed using the standard formalism \cite{zhang1998defect, Freysoldt2014}. As an example, the formation energy of sulfur vacancy ($V_\mathrm{S}$) in charge state $q$, i.e.,  $V_\mathrm{S}^q$, is given by:
\begin{equation*}
    E_\mathrm{f}(V_\mathrm{S}^q)=E_\mathrm{tot}(V_\mathrm{S}^q) - E_\mathrm{tot}(\mathrm{bulk}) + \mu_\mathrm{S} + qE_\mathrm{F} + \Delta^q,
\end{equation*}
where $E_\mathrm{tot}(V_\mathrm{S}^q)$ and $E_\mathrm{tot}(\mathrm{bulk})$ are the total energies of the defect-containing and defect-free supercells, respectively. The $\mu_\mathrm{S}$ is the sulfur chemical potential which can be understood as representing the sulfur content in the growth or annealing environment. Fig. S1 in the Supplemental Material \cite{supp} shows the chemical-potential stable region of BaZrS$_3$, which defines the allowed chemical-potential values for Ba, Zr, and S. Low $\mu_\mathrm{S}$ values means S-poor growth conditions thus lowering the formation energy of sulfur vacancies, and vice versa. The electron Fermi level $E_\mathrm{F}$ is referenced to the bulk valence-band maximum (VBM), and its position can be adjusted from the VBM to the conduction-band minimum (CBM). The term $\Delta^q$ is the finite-supercell-size correction for charged defects. We obtained charge corrections using the extended Freysoldt-Neugebauer-VandeWalle (FNV) scheme \cite{freysoldt2009fully,kumagai2014electrostatics} and the static dielectric constants ($\varepsilon_{xx}=82.10,~ \varepsilon_{yy}=70.67,~\varepsilon_{zz}=85.46$, which include both electronic and ionic contributions calculated using the PBE functional \cite{perdew1996generalized}).

To correctly find the ground-state configuration of sulfur interstitial (S$_i$), besides using the interstitial search algorithm implemented in the PyCDT code \cite{broberg2018pycdt} to generate 9 trial interstitial sites, we additionally generated another 13 trial interstitial sites based purely on random numbers (though these random interstitial sites are required to have reasonable distances to the lattice sites). 

Besides the isolated defects, the charge-compensated Frenkel pair $[V_\mathrm{S}$+S$_i]$ was also studied. Similar to the search for the ground-state configuration of S$_i$, we generated 13 trial configurations for the $[V_\mathrm{S}$+S$_i]$ pair. The binding energy ($E_\mathbf{b}$) of this defect pair was obtained as:
\begin{equation*}
    E_\mathbf{b}=E_\mathrm{f}(V_\mathrm{S}^{2+})+E_\mathrm{f}(\mathrm{S}_i^{2-})-E_\mathrm{f}([V_\mathrm{S}+\mathrm{S}_i]).
\end{equation*}
A positive binding energy means energy gain for the formation of the defect pair, but entropy favors the isolated constituents \cite{zhang1998defect,van2004first, Freysoldt2014}.

In thermodynamic equilibrium, the concentration of a defect in the charge state $q$ (denoted as $\mathrm{D}^q$) can be determined using its calculated formation energy,
\begin{equation*}
    C[\mathrm{D}^q] = N_\mathrm{sites} e^{-\frac{E_\mathrm{f}(\mathrm{D}^q)}{k_\mathrm{B}T}},
    \label{defectdensity}
\end{equation*}
where $N_\mathrm{sites}$ is the number of sites (per unit volume) at which the defect can form, $k_\mathrm{B}$ the Boltzmann constant, and $T$ the temperature. The equilibrium Fermi level can be determined by the charge-neutrality condition which requires the charges of positively charged defects and hole carriers balance the charges of negatively charged defects and electron carriers in the material \cite{van2004first,ma2011carrier,kumagai2014first,yang2014tuning}. This leads to a self-consistent solution of defect densities, carrier concentrations, and Fermi-level position. Often, materials are grown at high temperatures and measured at a lower temperature. To account for this, the defect densities can be computed at high temperatures, and one assumes that for each of the defect species, the total density is fixed to that at high temperatures but the charges are allowed to re-equilibrate so as to reach the charge-neutrality condition at a lower temperature. This is the so-called a rapid quenching process \cite{ma2011carrier,kumagai2014first,yang2014tuning}.

Several Python toolkits including the PyCDT \cite{broberg2018pycdt}, pymatgen \cite{ong2013python}, Pydefect \cite{kumagai2021insights}, and py-sc-fermi \cite{squires2023py} were used for generating initial defect structures, plotting chemical-potential phase diagram, and obtaining defect formation energies and densities. We also mention here the computer program VESTA \cite{momma2011vesta} was used for visualizing atomic geometries and the Python toolkit sumo \cite{ganose2018sumo} was used for plotting the electronic band structure.

\subsection{Nonradiative carrier capture calculations\label{section_mc}}

The calculation of nonradiative carrier capture via multiphonon emission was based on a quantum mechanical 1D model \cite{stoneham2001theory,lang1975nonradiative,henry1977nonradiative,huang1981lattice,alkauskas2014first}. In this 1D model, a generalized configuration coordinate ($Q$) is defined:
\begin{equation*}
    Q^2 = \sum_I M_I (\Delta\mathbf{R}_I)^2,
    \label{eqs2}
\end{equation*}
where $\Delta\mathbf{R}_I$ is the displacement of the $I$-th atom of mass $M_I$ along the vector connecting the positions of this atom in the equilibrium defect structures of the 
initial ($i$) and final ($f$) charge states. As such, there is only one (effective) vibrational degree of freedom. For both charge states, obtaining a set of intermediate configurations by a linear interpolation or extrapolation between the two equilibrium structures and computing their total energies lead to two potential energy surfaces. For each intermediate configuration, the electronic eigenstates were carefully checked to make sure that the defect states do not cross the band edges or undergo a change in occupation \cite{kim2019anharmonic, whalley2023, wang2024upper}. The two potential energy surfaces are offset horizontally (i.e., along the $Q$ axis) by $\Delta Q$, where $Q=\Delta Q$ and $Q=0$ correspond to the the equilibrium configurations of the initial and final charge states, and offset vertically (i.e., along the energy axis) by $\Delta E$, where $\Delta E$ is the charge transition level referenced to the VBM (CBM) for hole (electron) capture. A third potential energy surface, which is the potential energy surface of the final charge state but shifted vertically by the band gap value, is also needed to represent a free hole at the valence-band edge and a free electron at the conduction-band edge. With these, a configuration coordinate diagram is constructed for a carrier capture cycle (i.e., a complete process of nonradiative electron-hole recombination) \cite{lang1975nonradiative,henry1977nonradiative}. From the configuration coordinate diagram, one can analyze the electron and hole capture barriers which provide a qualitative estimate how fast the capture processes are.

The carrier capture coefficients were computed based on the Fermi's golden rule in the static coupling theory \cite{alkauskas2014first,Shi2015}:
\begin{equation*}
\begin{split}
    C(T) = &gV\frac{2\pi}{\hbar} \sum_m w_m \sum_n |\bra{\chi_{im}}Q-Q_0\ket{\chi_{fn}}|^2|W|^2\\
    &\times \delta(E_{im} - E_{fn}),
    \label{eqs3}
\end{split}
\end{equation*}
which requires computing the overlap of vibrational wave functions and the electron–phonon coupling strength. In this equation, $\chi_{im}$ ($\chi_{fn}$) are vibrational wave functions of the potential energy surface for the initial (final) charge state, and are specified by a set of occupation numbers $\{m\}$ ($\{n\}$). $w_m$ is the thermal weight. $W=\bra{\psi_\mathrm{b}}\partial h/\partial Q\ket{\psi_\mathrm{d}}$ is the electron-phonon matrix element in the first-order perturbation theory, where $\psi_\mathrm{b}$ and $\psi_\mathrm{d}$ are bulk and defect states for the single-particle electronic Hamiltonian $h$. The electron-phonon matrix element is evaluated at $Q_0$, a chosen configuration of which a well localized defect state can be identified in the band gap \cite{alkauskas2014first}. $g$ is the degeneracy factor of the final charge state. The delta function ensures energy conservation, where $E_{im}$ and $E_{fn}$ are total energies; in the harmonic approximation, $E_{im}-E_{fn}=\Delta E+m\hbar\omega_i-n\hbar\omega_f$, where $\omega_i$ and $\omega_f$ are phonon frequencies. We calculated the $W$ elements within the PAW formalism using VASP \cite{turiansky2021nonrad}. For carrier capture by a charged defect, $|W|$ was scaled by a Sommerfeld parameter ($s$) which accounts for the Coulomb attraction (or repulsion) between the delocalized carrier and charged defect; An additional scaling factor ($f$) was included for correcting the charge density near the defect when the $W$ elements are computed using a charged defect supercell. These scaling factors were computed using the Nonrad code \cite{turiansky2021nonrad}. The overlap of vibrational wave functions and final capture coefficients were determined using the CarrierCapture.jl code \cite{kim2019anharmonic,kim2020carriercapture}. We performed carrier capture calculations for the deep defect S$_i$, with the key material parameters detailed in Table S5 of the Supplemental Material \cite{supp}.

\section{Data Availability}

Input and output files for first-principles defect calculations in this work with relaxed defect structures are openly available on the NOMAD Repository \cite{nomad}, where files for the chemical-potential phase diagram, density of states, and (static) dielectric constant calculations are also available. Other data are available upon reasonable request from the authors.

\begin{acknowledgments}
This work was supported by the U.S. Department of Energy, Office of Science, Basic Energy Sciences under award number DE-SC0023509. This research used resources of the National Energy Research Scientific Computing Center (NERSC), a DOE Office of Science User Facility supported by the Office of Science of the U.S. Department of Energy under contract no. DE-AC02-05CH11231 using NERSC award BES-ERCAP0023830. A.P. acknowledges support from a Department of Education GAANN fellowship. 
\end{acknowledgments}

\bibliography{Refs}

\end{document}